\documentstyle[amssymb,preprint,aps]{revtex}
\tightenlines

\begin{document}
\title{de Broglie's pilot wave theory for the Klein-Gordon equation}
\author{George Horton, Chris Dewdney and Ulrike Neeman}

\begin{abstract}
We illustrate, using a simple model, that in the usual formulation the
time-component of the Klein-Gordon current is not generally positive
definite even if one restricts allowed solutions to those with positive
frequencies. Since in de Broglie's theory of particle trajectories the
particle follows the current this leads to difficulties of interpretation,
with the appearance of trajectories which are closed loops in space-time and
velocities not limited from above. We show that at least this pathology can
be avoided if one uses a covariant extension of the canonical formulation of
relativistic point particle dynamics proposed by Gitman and Tyutin.
\end{abstract}

\maketitle

\section{Particle trajectory interpretations of the Klein-Gordon equation}

de Broglie first proposed a particle-trajectory interpretation of the
Klein-Gordon equation in the period 1926-1927\cite{deBroglie27}. He
proceeded using the polar form of the scalar field, 
\begin{equation}
\phi =R\exp (iS)
\end{equation}
to decompose the wave equation (with $\hbar =c=1$) 
\begin{equation}
\left( \partial ^{\mu }+ieA^{\mu }\right) \left( \partial _{\mu }+ieA_{\mu
}\right) \phi =-m_{0}^{2}\phi
\end{equation}
into a continuity equation 
\begin{equation}
\partial ^{\mu }(R^{2}\left( \partial _{\mu }S+eA^{\mu }\right) )=0
\label{eq:deBrogliec}
\end{equation}
and a ``Hamilton-Jacobi'' equation 
\begin{equation}
\left( \partial _{\mu }S+eA_{\mu }\right) \left( \partial ^{\mu }S+eA^{\mu
}\right) -\frac{\square R}{R}=m_{0}^{2}
\end{equation}
thus casting the theory in the form of a ``classical-particle'' theory with
canonical four-momentum $\partial ^{\mu }S$ and an additional ``quantum
potential'' term $-\frac{\square R}{R}$ which de Broglie interpreted in
terms of a variable rest mass $\sqrt{m_{0}^{2}+\frac{\square R}{R}}$.
Particles follow the streamlines determined by equation \ref{eq:deBrogliec}
with three-velocity defined by de Broglie as 
\begin{equation}
v_{dB}^{k}=\frac{S^{k}+eA^{k}}{-\left( S^{0}+eA^{0}\right) }  \label{debvel}
\end{equation}
with $k=1,2,3$ and where $S^{\mu }=\partial ^{\mu }S$ and so on. However,
the problem with the use of $\partial _{\mu }S$ to define the flow lines,
and hence particle trajectories, is that $\partial _{\mu }S$ is not always a
time-like four vector. Furthermore, as we demonstrate in a specific case
below, the time component of the current, $R^{2}\partial _{0}S,$ is not
always positive definite even if one restricts the allowed solutions to
superpositions of those with positive frequencies only\cite{Gerlach67} (see
also \cite{Prugovecki94}). The above is often taken to imply the breakdown
of the single-particle interpretation of the Klein-Gordon equation and to
signal the unavoidable effects of negative energies and hence the need for
anti-particles in the description.

\subsection{Appearance of a negative time-component of the current}

A general proof of the existence of a non positive-definite time-component
of the current has been given in \cite{Gerlach67} and \cite{Prugovecki94}.
Here we illustrate this point by considering the following simple
superposition of two positive frequency eigenstates of the Klein-Gordon
equation 
\begin{equation}
\Phi \left( x,t\right) =\phi _{1}\left( x\right) e^{-i\omega _{1}t}+\phi
_{2}\left( x\right) e^{-i\omega _{2}t}
\end{equation}
and let us assume that $\phi _{1}$and $\phi _{2}$ are real functions of $x$
and that $\omega _{1}$and $\omega _{2}$ are positive. The time-component of
the current is given by 
\begin{eqnarray}
J^{0} &=&-\left| \Phi \right| ^{2}\frac{\hbar }{m_{0}}%
\mathop{\rm Im}%
\left[ \frac{\frac{\partial \Phi }{\partial t}}{\Phi }\right] \\
&=&\frac{\hbar }{m_{0}}\left[ \omega _{1}\left| \phi _{1}\right| ^{2}+\omega
_{2}\left| \phi _{2}\right| ^{2}+\phi _{1}\phi _{2}\left( \omega _{1}+\omega
_{2}\right) \cos \left( \left( \omega _{1}+\omega _{2}\right) t\right) %
\right]
\end{eqnarray}
and this may be backwards pointing in time if 
\begin{equation}
\cos \left( \left( \omega _{1}+\omega _{2}\right) t\right) <-\left[ \frac{%
\omega _{1}\left| \phi _{1}\right| ^{2}+\omega _{2}\left| \phi _{2}\right|
^{2}}{\phi _{1}\phi _{2}\left( \omega _{1}+\omega _{2}\right) }\right]
\end{equation}
and since $\phi _{1}\phi _{2}$ can be negative the inequality can easily be
satisfied.

\subsubsection{A specific example: the scalar potential box.}

Although electromagnetic potentials are incapable of confining a
relativistic particle, scalar potentials can do so. Using such a potential
one can create a one-dimensional scalar potential ``box'' of length $L$ such
that $V=0$ for $0<x<L$ and $V=\infty $ otherwise. The eigenstates are given
by 
\begin{equation}
\phi _{n}=\sqrt{\frac{2}{L}}\sin \left( \frac{n\pi x}{L}\right) \exp \left(
-i\omega _{n}t\right)
\end{equation}
where $n$ is an integer and 
\begin{equation}
\omega _{n}=c\sqrt{\left( \frac{n\pi }{L}\right) ^{2}+\left( \frac{m_{0}c}{%
\hbar }\right) ^{2}}
\end{equation}
A simple superposition of states serves to illustrate the pathologies of de
Broglie's formulation. We consider the state 
\begin{equation}
\Phi \left( x,t\right) =\frac{1}{\sqrt{2}}\left[ \phi _{1}+\phi _{2}\right]
\label{eq:super}
\end{equation}
and for the purposes of the calculation we take $L=\pi $ and $\hbar
=c=m_{0}=1.$

In figure 1 we plot $\left| \Phi \right| ^{2}$ (dotted line), the de Broglie
three-velocity, given by equation \ref{debvel}, (solid line) and the time
component of the four velocity $S_{0}$ (dashed line) for the state given in
equation \ref{eq:super} at $t=0.1.$ Clearly there are regions in which the
three-velocity becomes superluminal (this is where the three velocity has a
magnitude greater than unity in our chosen units). The de Broglie
three-velocity is discontinuous where $S_{0}$ changes sign, for the values
of the parameters chosen for the plot this occurs at $x\simeq 2.086$ and $%
x\simeq 1.898$ (to three decimal places). Figure 2 shows the space-time
flows for this simple example. The flows are calculated from the
four-velocity with 
\begin{eqnarray}
\frac{\partial t\left( \tau \right) }{\partial \tau } &=&-S^{0}
\label{eq:4vdB} \\
\frac{\partial x\left( \tau \right) }{\partial \tau } &=&S^{1}
\end{eqnarray}
One has closed loops in space-time which are difficult to interpret. As far
as we know de Broglie never commented on this pathological situation. To
consider this point more carefully one has to consider the meaning of Gauss'
theorem in four-dimensional space-time.

\section{Application of Gauss' theorem to time-like and space-like flows}

Gauss' theorem in four-dimensional space-time is 
\begin{equation}
\int\limits_{V^{4}}\partial _{\mu }\left( u^{\mu }\right)
d^{4}x=\int\limits_{V^{3}}\left( u^{\mu }\right) \epsilon \left( n\right)
n_{\mu }d^{3}x
\end{equation}
where $\epsilon \left( n\right) =\pm 1$ depending on the time-like or
space-like character of $n_{\mu }$ \cite{synge}. One can interpret this in
terms of time-like or space-like conserved flows of $u^{\mu }$ through
hypersurfaces. For example, one can consider flow through a world-tube
bounded by the hypersurface parallel to the flow and hypersurfaces normal to
the flow at each end. (If the flow crosses a null hypersurface a limiting
process must be used.)

Consider the application of Gauss' theorem to the example previously
discussed and illustrated in figure 2 where there are closed paths in
space-time. In order to discuss the slicing of the flow tube normal to the
direction of flow we consider the Fermi transport of an orthogonal dyad (in
general a tetrad) of a time-like and a space-like vector around the closed
loop. This provides a natural way of defining the orientation of the
surfaces with respect to the flow as illustrated in figure 3. One should
note that the orthogonal vectors maintain their orientation with respect to
the time and the space axes. The full details of Fermi transport along
time-like and space-like curves are given in reference \cite{synge}. An
application of Gauss' theorem now gives a conserved flow around the closed
path in space-time. There remains the interpretative difficulty associated
with backward-pointing time-like flows in regions where this is an invariant
characterisation (e.g. the region of figure 3 bounded by the light-cone and
containing the positive portion of the x-axis). In all other regions the
space-like flows do not have an invariant characterisation in terms of
orientation in time.

It seems, however, that a more careful formulation of the dynamics allows
one to remove at least the pathology of the backward-pointing time-like
flows, as we now proceed to demonstrate.

\section{Classical relativistic dynamics revisited}

In the foregoing we have followed de Broglie in formulating the theory in a
particular reference frame. However, following the non-covariant treatment
ofGitman and Tyutin \cite{Gitman1990}, and \cite{hannibal1991}, the de
Broglie theory can be cast in covariant form in the following way.

The classical action for a relativistic, spinless particle can be written as 
\begin{equation}
S=-m\int Ld\tau
\end{equation}
where 
\begin{equation}
L=-m\sqrt{\dot{x}_{\mu }\dot{x}^{\mu }}
\end{equation}
and $\dot{x}_{\mu }=\frac{dx_{\mu }}{d\tau }$with $\tau $ a scalar parameter
along the path of the particle. The canonical momenta $\pi _{\mu }$are then
given by 
\begin{equation}
\pi _{\mu }=\frac{\partial L}{\partial \dot{x}^{\mu }}=-m\frac{\dot{x}_{\mu }%
}{\sqrt{\dot{x}_{\mu }\dot{x}^{\mu }}}  \label{eq:canon}
\end{equation}
from which one has the primary constraint 
\begin{equation}
\pi _{\mu }\pi ^{\mu }=m^{2}
\end{equation}
Decomposing $\pi ^{\mu }$ along a unit time-like four-vector $n^{\mu }$ $%
\left( n^{0}>0\right) $ and perpendicular to it one has 
\begin{equation}
\pi ^{\mu }=\hat{\pi}^{\mu }+\left( \pi _{\nu }n^{\nu }\right) n^{\mu }
\end{equation}
where $\hat{\pi}^{\mu }$is space-like and $\hat{\pi}^{\mu }\hat{\pi}_{\mu
}<0 $. (In the following the same covariant decomposition notation will be
used, the caret accent indicating the space-like part of the vector.)
Therefore 
\begin{equation}
\pi _{\mu }\pi ^{\mu }=\hat{\pi}^{\mu }\hat{\pi}_{\mu }+\left( \pi _{\nu
}n^{\nu }\right) ^{2}=m^{2}
\end{equation}
Putting $\pi _{\nu }n^{\nu }=\Pi ,$ the primary constraint becomes 
\begin{equation}
\Phi =\sqrt{-\hat{\pi}_{\mu }\hat{\pi}^{\mu }+m^{2}}-\left| \Pi \right| =0
\end{equation}
which is a covariant formulation of the constraint. Resolving both sides of
equation \ref{eq:canon} along along $n^{\mu }$ and perpendicular to it one
has 
\begin{eqnarray}
\hat{\pi}_{\mu } &=&\frac{-m}{\sqrt{\dot{x}_{\nu }\dot{x}^{\nu }}}\widehat{%
\dot{x}}_{\mu } \\
\Pi &=&\frac{-m}{\sqrt{\dot{x}_{\nu }\dot{x}^{\nu }}}\dot{X}
\end{eqnarray}
where $\dot{X}=\left( \dot{x}_{\mu }n^{\mu }\right) $ and both $\Pi $ and $%
\dot{X}$ are scalars. Therefore 
\begin{equation}
\widehat{\dot{x}}_{\mu }=\hat{\pi}_{\mu }\frac{\dot{X}}{\Pi }
\end{equation}
and since $\Pi $ and $\dot{X}$ are of opposite sign, 
\begin{equation}
\widehat{\dot{x}}_{\mu }=-\hat{\pi}_{\mu }\frac{\left| \dot{X}\right| }{%
\left| \Pi \right| }=\frac{-\hat{\pi}_{\mu }}{\sqrt{-\hat{\pi}_{\mu }\hat{\pi%
}^{\mu }+m^{2}}}\left| \dot{X}\right|
\end{equation}
The Hamiltonian $H$ is given, as usual, by 
\begin{equation}
H=\pi _{\mu }\dot{x}^{\mu }-L
\end{equation}
which vanishes along the constraint surface. Following \cite{Gitman1990} one
can choose a special gauge (in a covariant way) as 
\begin{equation}
\Phi ^{\left( 0\right) }=X-\zeta \tau =0
\end{equation}
One can now switch to canonical variables $x^{\mu \prime },\pi ^{\mu \prime
} $using a generating function $W$ where 
\begin{eqnarray}
W &=&x^{\mu }\pi _{\mu }^{\prime }+\tau \left| \Pi \right| \\
X^{\prime } &=&X-\zeta \tau \\
x_{\mu }^{\prime } &=&x_{\mu } \\
\pi _{\mu } &=&\pi _{\mu }^{\prime }
\end{eqnarray}
The new Hamiltonian $H^{\prime }$ is 
\begin{equation}
H^{\prime }=H+\frac{\partial W}{\partial \tau }=\left| \Pi \right|
\end{equation}
on the constraint surface. Therefore 
\begin{equation}
H^{\prime }=\sqrt{-\hat{\pi}_{\mu }\hat{\pi}^{\mu }+m^{2}}
\end{equation}
and since $\dot{X}=\zeta $%
\begin{equation}
\widehat{\dot{x}}_{\mu }=\frac{-\hat{\pi}_{\mu }}{\sqrt{-\hat{\pi}_{\mu }%
\hat{\pi}^{\mu }+m^{2}}}\left| \zeta \right|
\end{equation}
As in \cite{Gitman1990}, $\zeta $is not fixed by the constraints and can be
considered as another dynamical variable taking values $\pm 1$. In \cite
{Gitman1990} it is shown that the two values of $\zeta $ correspond to
particle and antiparticle motion.

Introducing an external electromagnetic field $A_{\mu }$ one obtains as
usual 
\begin{eqnarray}
H &=&\left( -\left( \hat{\pi}_{\mu }+e\hat{A}_{\mu }\right) \left( \hat{\pi}%
^{\mu }+e\hat{A}^{\mu }\right) +m^{2}\right) ^{\frac{1}{2}} \\
\frac{d\widehat{x}^{\mu }}{d\tau } &=&-\frac{\left( \hat{\pi}^{\mu }+e\hat{A}%
^{\mu }\right) }{H} \\
\frac{d\pi ^{\mu }}{d\tau } &=&-\frac{e\hat{A}^{\nu ,\mu }\left( \hat{\pi}%
_{\nu }+e\hat{A}_{\nu }\right) }{H}
\end{eqnarray}
Calling $X_{0}=\zeta \tau $ and $\wp ^{\mu }=\zeta \pi _{\mu }$ physical
time and momenta, respectively one finds 
\begin{eqnarray}
\frac{d\hat{x}^{\mu }}{dX} &=&\frac{\hat{\wp}^{\mu }-g\hat{A}^{\mu }}{\left(
-\left( \hat{\wp}^{\mu }-g\hat{A}^{\mu }\right) ^{2}+m^{2}\right) ^{\frac{1}{%
2}}} \\
\frac{d\hat{\wp}^{\mu }}{dX} &=&\frac{-g\hat{A}^{\nu ,\mu }\left( \hat{\wp}%
^{\mu }-g\hat{A}^{\mu }\right) }{\left( -\left( \hat{\wp}^{\mu }-g\hat{A}%
^{\mu }\right) ^{2}+m^{2}\right) ^{\frac{1}{2}}}
\end{eqnarray}
where $g=\zeta e$. Therefore trajectories with $\zeta =1$ correspond to a
particle with charge $e$ and those with $\zeta =-1$ to a particle with
charge $-e.$ One also notes that, since $X=x_{\mu }n^{\mu }$ with $n^{0}>0$,
the scalar time parameter $X$ is positive for both cases (or negative for
both cases).

In a given frame with $n^{\mu }$ coincident with the time axis $\Pi =\pi
_{0} $. So following this approach in de Broglie's particle theory,
remembering the connection of $S^{\mu }$ with the canonical momenta one has
to take the magnitude of $S_{0}$ in calculating the three-velocity;
yielding, rather than equation \ref{debvel}, in a given frame and in the
absence of potentials 
\begin{equation}
v_{dB}^{k}=\frac{S^{k}}{\left| S^{0}\right| }  \label{eq:vdBmod}
\end{equation}
For the particular case, discussed above, of the square well with the
wavefunction $\Phi $ defined by \ref{eq:super}, this modified form of the
velocity is plotted in figure 4 as the solid line, along with $\left| \Phi
\right| ^{2},$ as the dotted line, for comparison. The two forms of the de
Broglie velocity coincide except where $S^{0}$is positive. Figure 5 shows
the space-time flows calculating using 
\begin{eqnarray}
\frac{\partial t\left( \tau \right) }{\partial \tau } &=&\left| S^{0}\right|
\label{eq:dB4dmod} \\
\frac{\partial x\left( \tau \right) }{\partial \tau } &=&S^{1}
\end{eqnarray}
which obviously do not have space-time loops, although the flows still
become space-like in certain regions.

\section{Time-like trajectories}

Elsewhere \cite{Dewdney96},\cite{Horton 2000} we have given details of an
approach to the definition of particle trajectories for the Klein-Gordon
equation which yields time-like trajectories in all circumstances. Our
Lorentz invariant description defines the flow of stress-energy-momentum,
and hence particle trajectories which follow the flow, through the intrinsic
natural four-vector provided by the matter field itself through the
eigenvalue equation 
\begin{equation}
T_{\nu }^{\mu }W^{\nu }=\lambda W^{\mu }  \label{eigenvalue}
\end{equation}
where $T_{\nu }^{\mu }$ is the stress-energy-momentum tensor, $\lambda $ the
eigenvalue and $W^{\mu }$ the eigenvector. Writing a solution $\phi $ of the
Klein-Gordon equation as 
\begin{equation}
\phi =\exp [P+iS]
\end{equation}
the stress-energy-momentum tensor, $T_{\nu }^{\mu },$ of the field $\phi $
is given by 
\begin{equation}
T_{\nu }^{\mu }=|\phi |^{2}[m_{0}^{2}-(P^{\alpha }P_{\alpha }+S^{\alpha
}S_{\alpha })]\delta _{\nu }^{\mu }+2|\phi |^{2}[(P^{\mu }P_{\nu }+S^{\mu
}S_{\nu })]
\end{equation}
Once the state $\phi $ is given the stress-energy-momentum tensor can be
calculated along with its eigenvalues and eigenvectors. As we showed in \cite
{Dewdney96}, one finds a pair of eigenvectors one of which is time-like and
the other space-like. The time-like vector and its eigenvalue determine the
flows of energy and the density respectively. Applying this approach in this
simple, one-dimensional example yields the one-dimensional three-velocity 
\begin{equation}
v^{1}=\frac{W^{1}}{W^{0}}=\frac{-\left( T_{00}-\lambda \right) }{T_{01}}
\label{eq:vhT}
\end{equation}
where 
\begin{equation}
\lambda =\frac{T_{00}-T_{11}}{2}\pm \sqrt{\frac{\left( T_{00}-T_{11}\right)
^{2}}{4}-\left( T_{01}\right) ^{2}}
\end{equation}
More generally (in more than one dimension) the velocity is given by 
\begin{equation}
v^{k}=\frac{S^{k}\pm e^{\pm \theta }\nabla P}{-\left( \frac{\partial S}{%
\partial t}\pm e^{\pm \theta }\frac{\partial P}{\partial t}\right) }
\label{eq:vh}
\end{equation}
where 
\begin{equation}
sinh\theta =\frac{P^{\mu }P_{\mu }-S^{\mu }S_{\mu }}{2P^{\mu }S_{\mu }}
\end{equation}
For our illustrative example, the three-velocity, given by equation \ref
{eq:vh} with the state \ref{eq:super} is plotted in figure 6, again for the
same parameter values used in the previous plots. The particle is moving
slowly throughout the whole region, with no extremes as exist in the de
Broglie velocities for this situation. Figure 7 shows the associated
trajectories. In the case of the unmodified de Broglie velocity the region
of this plot contained the space-time vortex, here we see perfectly regular
flow through this region governed by \ref{eq:vh}

\section{Conclusion}

We have shown that if one wishes to formulate an interpretation of the
Klein-Gordon equation based on individual particle trajectories then this
can be done in a consistent manner but only by going beyond the approach
originally proposed by de Broglie. In de Broglie's approach one has a non
positive-definite time-component of the current, closed loops in the
space-time flows, space-like particle motions and a rest mass which may
become imaginary. The situation can be improved by formulating the theory in
a covariant manner as one consequence of doing this properly is to remove
the possibility of backward-pointing time-like flows. However, this
modification of de Broglie's theory does not overcome the problems
associated with space-like trajectories. A particle-trajectory
interpretation which uses the flows defined by the
stress-stress-energy-momentum tensor of the Klein-Gordon field to determine
the particle trajectories does not suffer from the pathologies discussed in
de Broglie's theory. Adopting the latter approach enables a consistent
interpretation of the simple quantum phenomena associated with relativistic
particles in terms of well-defined particle trajectories. At least the
obstacles to the construction of a consistent particle interpretation of the
Klein-Gordon equation that we have discussed here can be removed, and a
consistent theory developed. The problems that we have discussed in this
paper are often cited as reasons for the rejection of a particle-trajectory
interpretation of the Klein-Gordon equation and the compulsion to adopt a
basic field ontology for bosons as suggested by Bohm. We have shown that if
one wishes to reject the particle ontology in favour of the field ontology
then the justification must be sought elsewhere.

\newpage {\bf Figure Captions}

\begin{enumerate}
\item  FIG.1. $\left| \Phi \right| ^{2}$ (dotted line), the de Broglie
three-velocity, given by equation \ref{debvel}, (solid line), the time
component of the four velocity $S_{0}$ (dashed line) plotted across the box,
for the state given by equation \ref{eq:super} with $L=\pi ,t=0.1.$

\item  FIG. 2. The flows in space-time determined from the de Broglie
equations \ref{eq:4vdB} and engendered by the state given in equation \ref
{eq:super} with $L=\pi $. The integration was peformed from the set of
initial conditions given by $\left( x_{0},\tau _{0}\right) =$ $\left[ \left(
1.9,-0.04\right) ,\left( 1.9,-0.1\right) ,\left( 2.4,-0.4\right) ,\left(
2.3,-0.4\right) ,\left( 2.0,-0.4\right) \right] $ chosen to display the
pathology around the minimum in $\left| \Phi \right| ^{2}$shown in figure 1.

\item  FIG. 3. Fermi transport of an orthogonal dyad around a closed path in
space-time. The solid arrow is the time-like vector and the outlined arrow
the space-like vector, together they form an orthogonal dyad. The time-like
vector maintains its orientation with respect to the {\it t} axis and
similarly the space-like vector with respect to the {\it x }axis.

\item  FIG. 4. $\left| \Phi \right| ^{2}$ (dotted line) and the modified
form of the de Broglie three-velocity, given by equation \ref{eq:vdBmod}
plotted across the box, for the state given by equation \ref{eq:super} with $%
L=\pi ,t=0.1.$ for comparison with FIG 1.

\item  FIG. 5. The flows in space-time determined from equations \ref
{eq:dB4dmod} and engendered by the state given by equation \ref{eq:super}
with $L=\pi $. The integration was peformed from the set of initial
conditions given by $\left( x_{0},\tau _{0}\right) =$ $\left[ \left(
1.9,-0.04\right) ,\left( 1.9,-0.1\right) ,\left( 2.4,-0.4\right) ,\left(
2.3,-0.4\right) ,\left( 2.0,-0.4\right) \right] $ as for figure 2.

\item  FIG. 6. The three-velocity associated with the time-like eigenvector
of the stress-energy-momentum tensor for the state given by equation \ref
{eq:super} with $L=\pi ,t=0.1.$ This velocity is perfectly regular, unlike
either of the de Broglie velocities defined in equations \ref{debvel} or \ref
{eq:vdBmod}.

\item  FIG. 7. The flow lines governed by equation \ref{eq:vh} derived from
the eigenvectors of the stress-stress-energy-momentum tensor. The
integration was peformed from the set of initial conditions given by $\left(
x_{0},\tau _{0}\right) =$ $\left[ \left( 1.94,-0.4\right) ,\left(
2.0,-0.4\right) ,\left( 2.14,-0.4\right) ,\left( 2.3,-0.4\right) ,\left(
2.4,-0.4\right) \right] $.
\end{enumerate}


\begin{references}
\bibitem{deBroglie27}  de Broglie L, 1926 {\it C.R. Acad. Sci.} {\bf 183}
p447; 1927{\bf 185}, p380.

\bibitem{Gerlach67}  Gerlach B, Gromes D, Petzold J, 1967 {\it J. Z.Phys} 
{\bf 202}, p401; {\bf 204} p1.

\bibitem{synge}  Synge J.L, 1964 {\it Relativity: The General Theory.
(Amsterdam: North Holland).}

\bibitem{Gitman1990}  Gitman D M and Tyutin I V, 1990 {\it Class. Quantum
Grav.} {\bf 7} p2131.

\bibitem{hannibal1991}  Hannibal L, 1991 {\it International Journal of
Theoretical Physics. }{\bf 30}, p1431.

\bibitem{Prugovecki94}  Prugovecki E, 1994 {\it Principles of Quantum
General Relativity}. (New Jersey:World Scientific )

\bibitem{Dewdney96}  Dewdney C and Horton G, 1996 {\it Bohmian Mechanics and
Quantum Theory: An appraisal.} edited by Cushing J T, Fine A and Goldstein S
(Netherlands: Kluwer Academic Publishers) p169.

\bibitem{Horton 2000}  Horton G and Dewdney C, 2000 {\it J. Phys. A.} {\bf 33}, p7337.

\end{references}
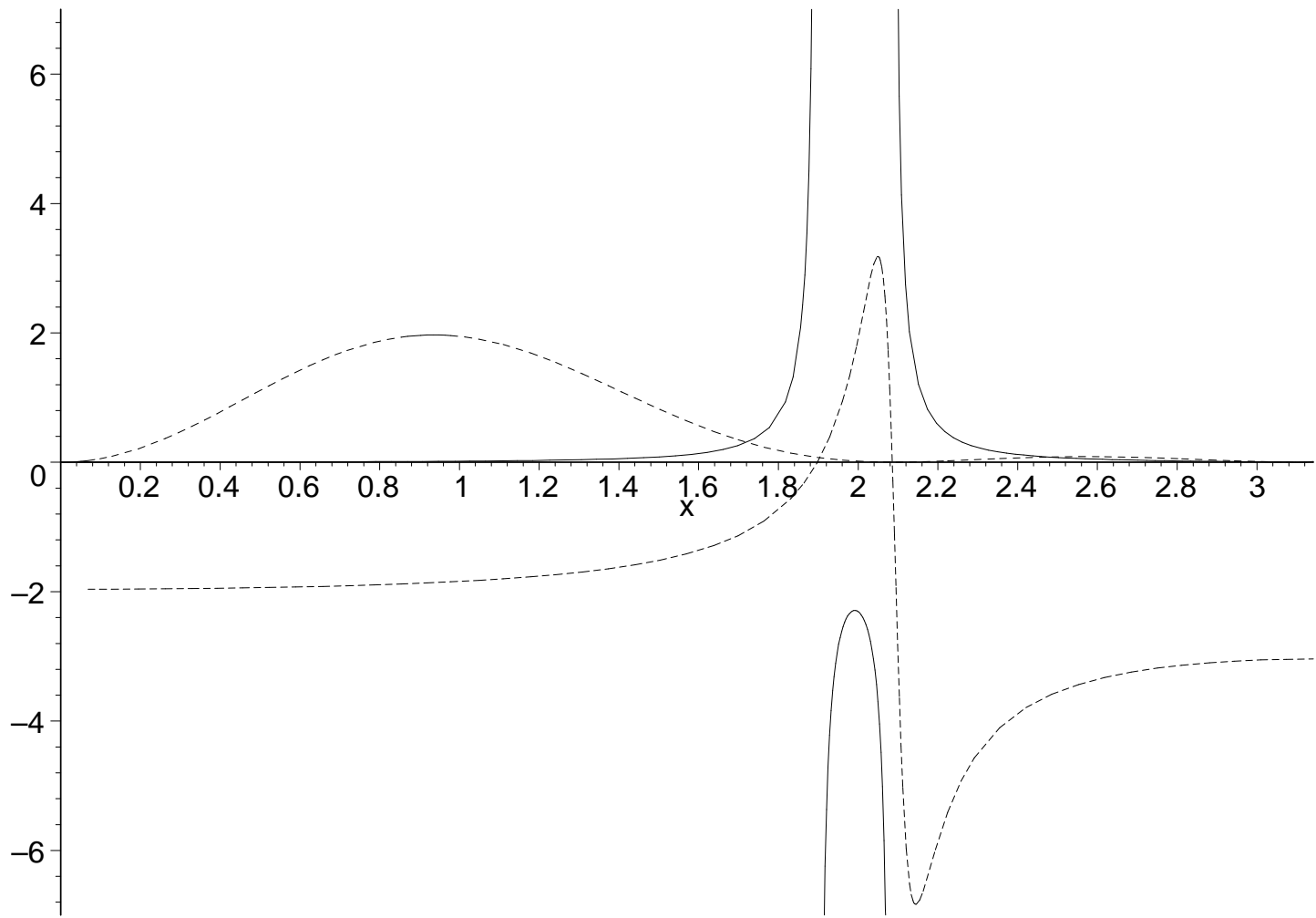
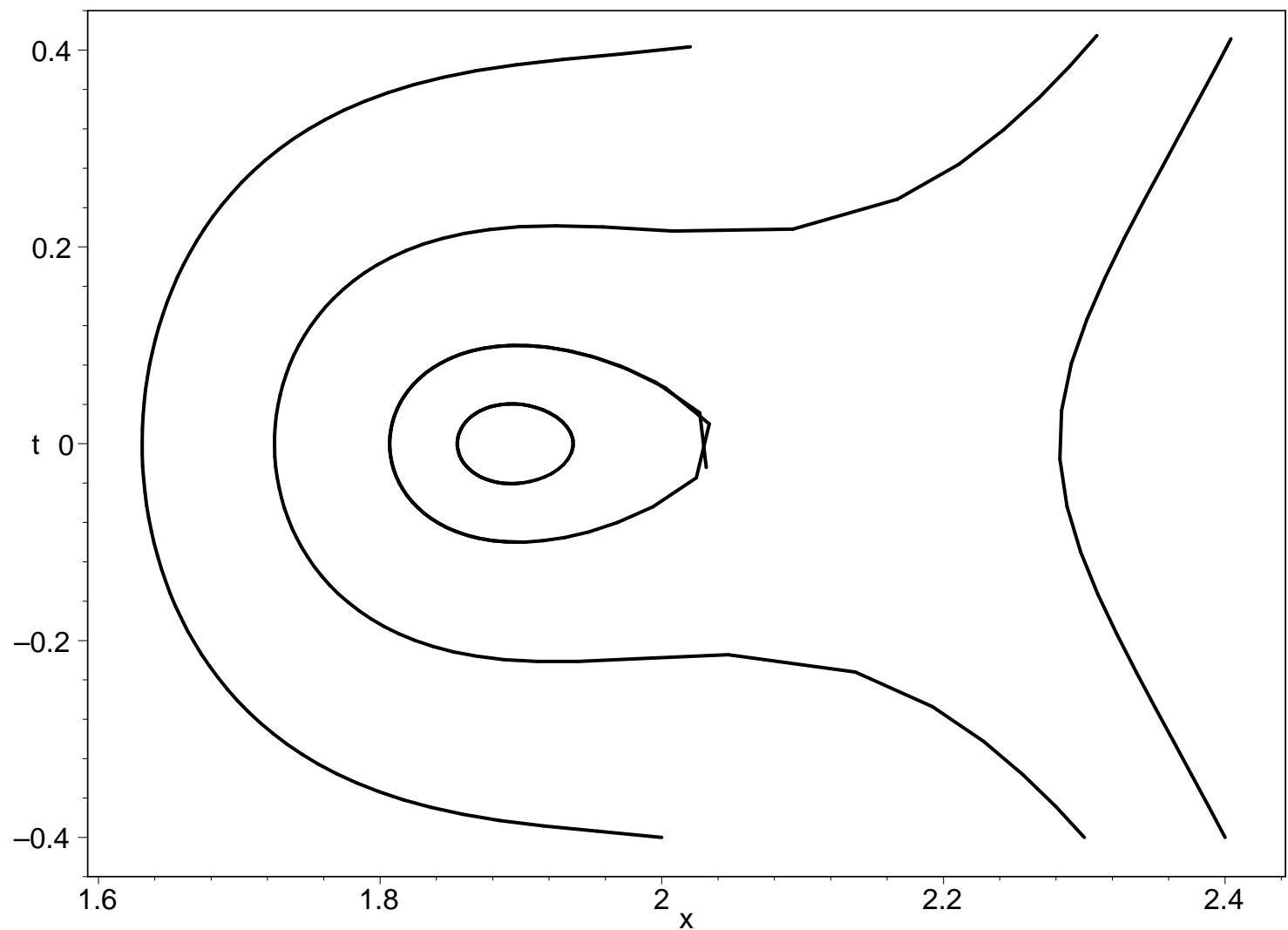
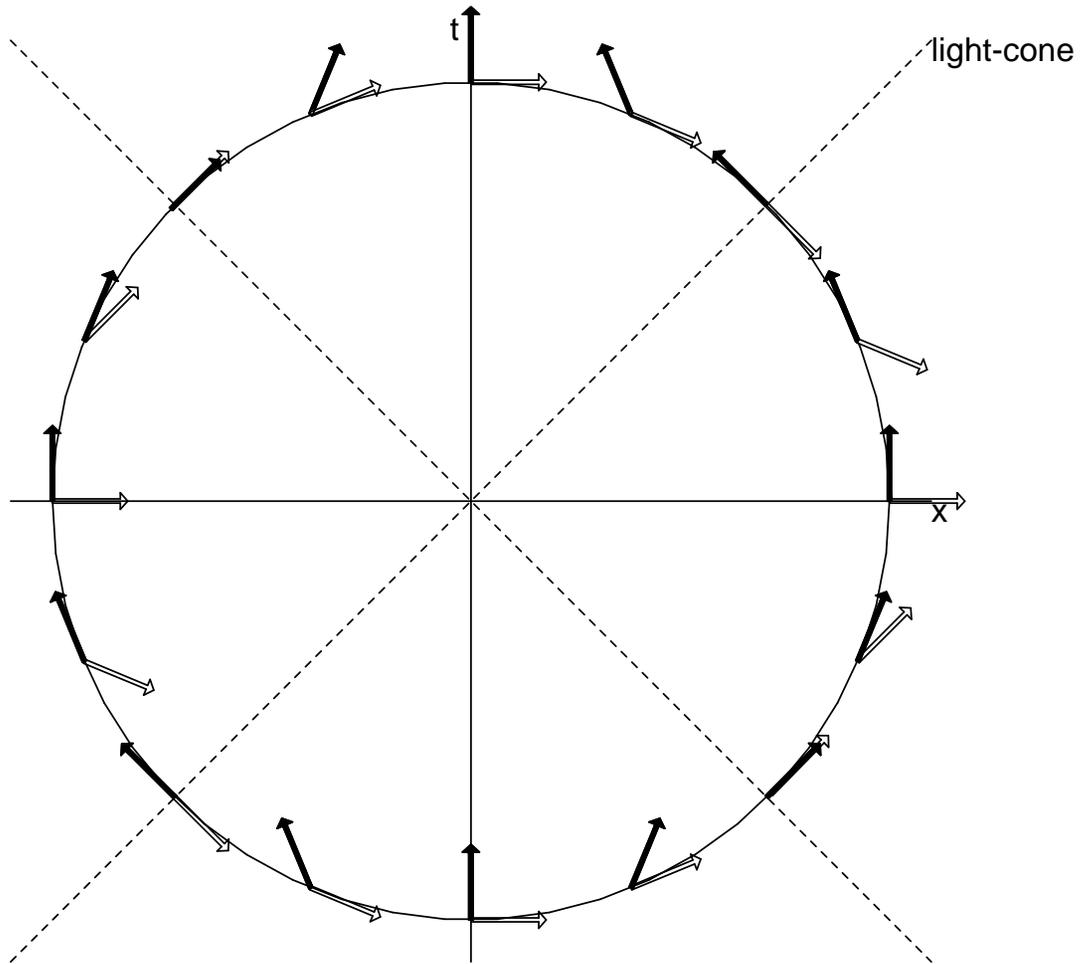
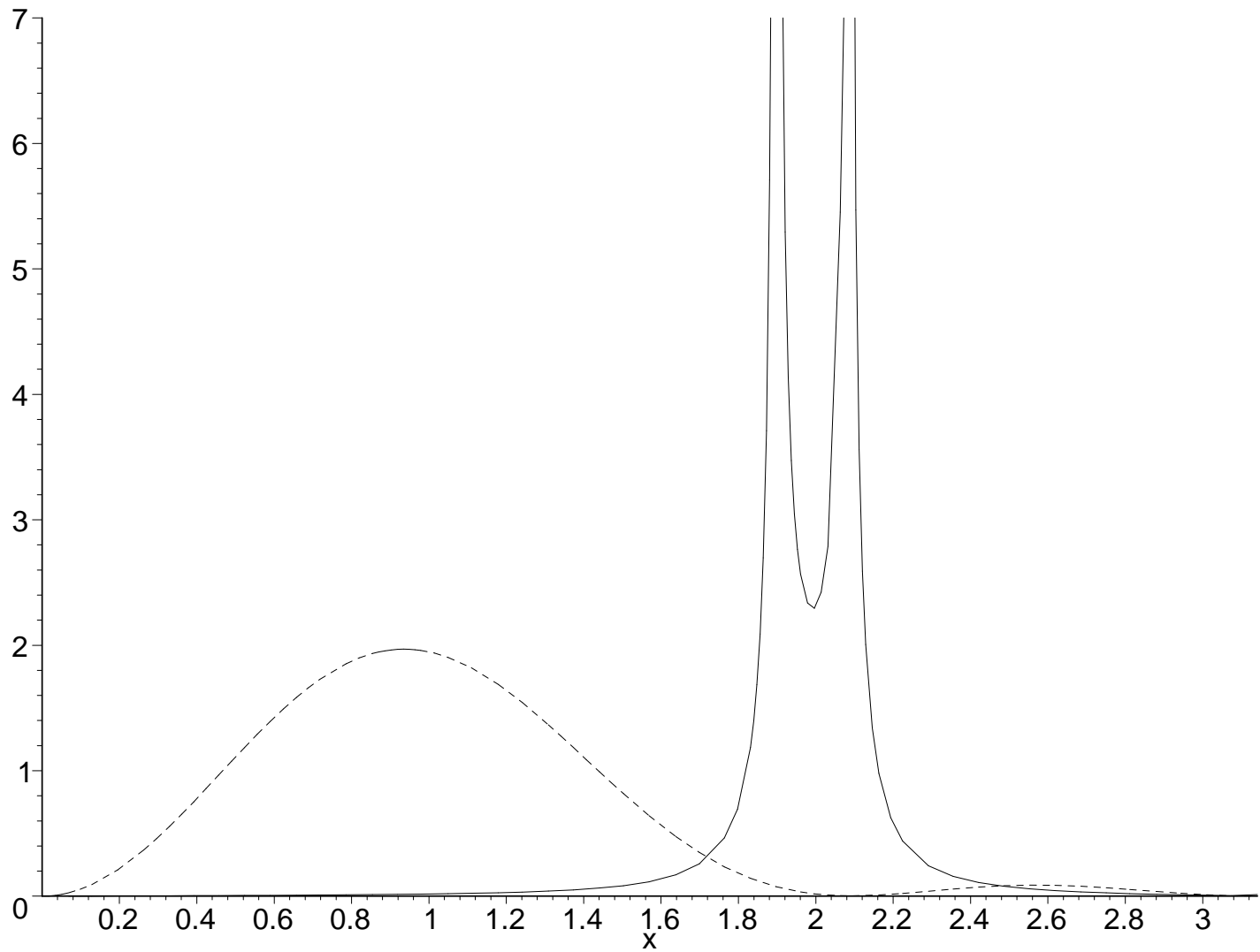
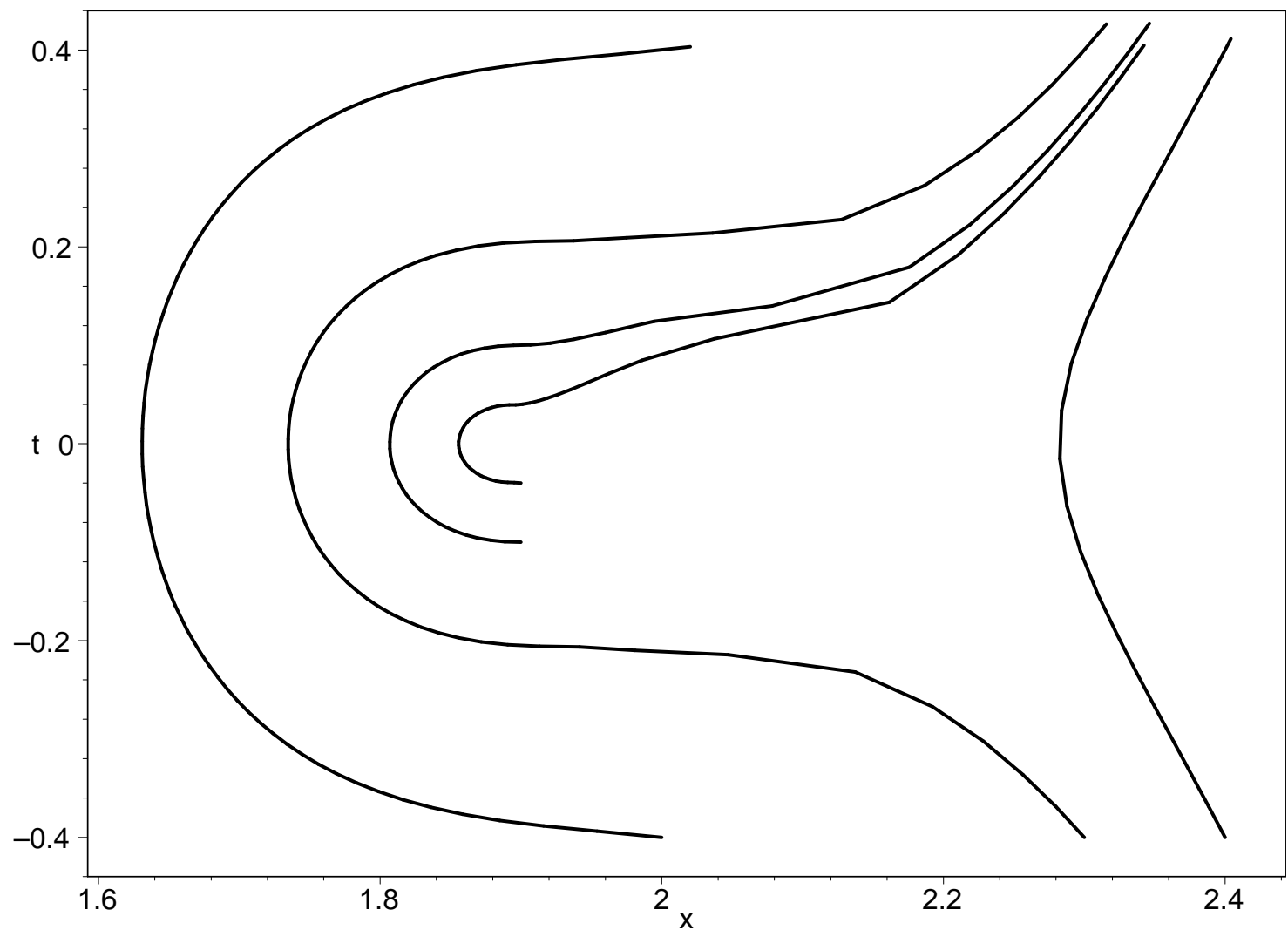
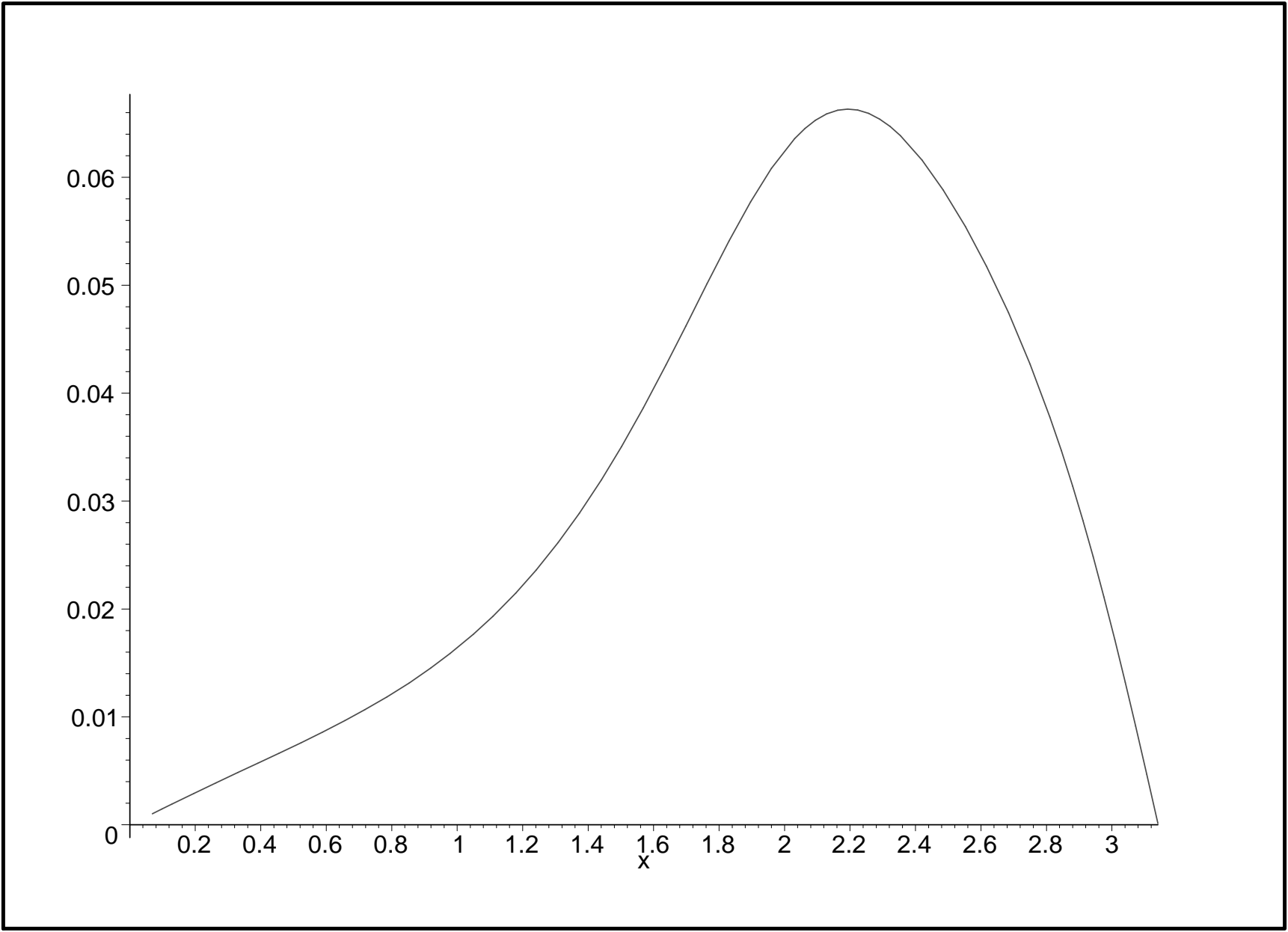
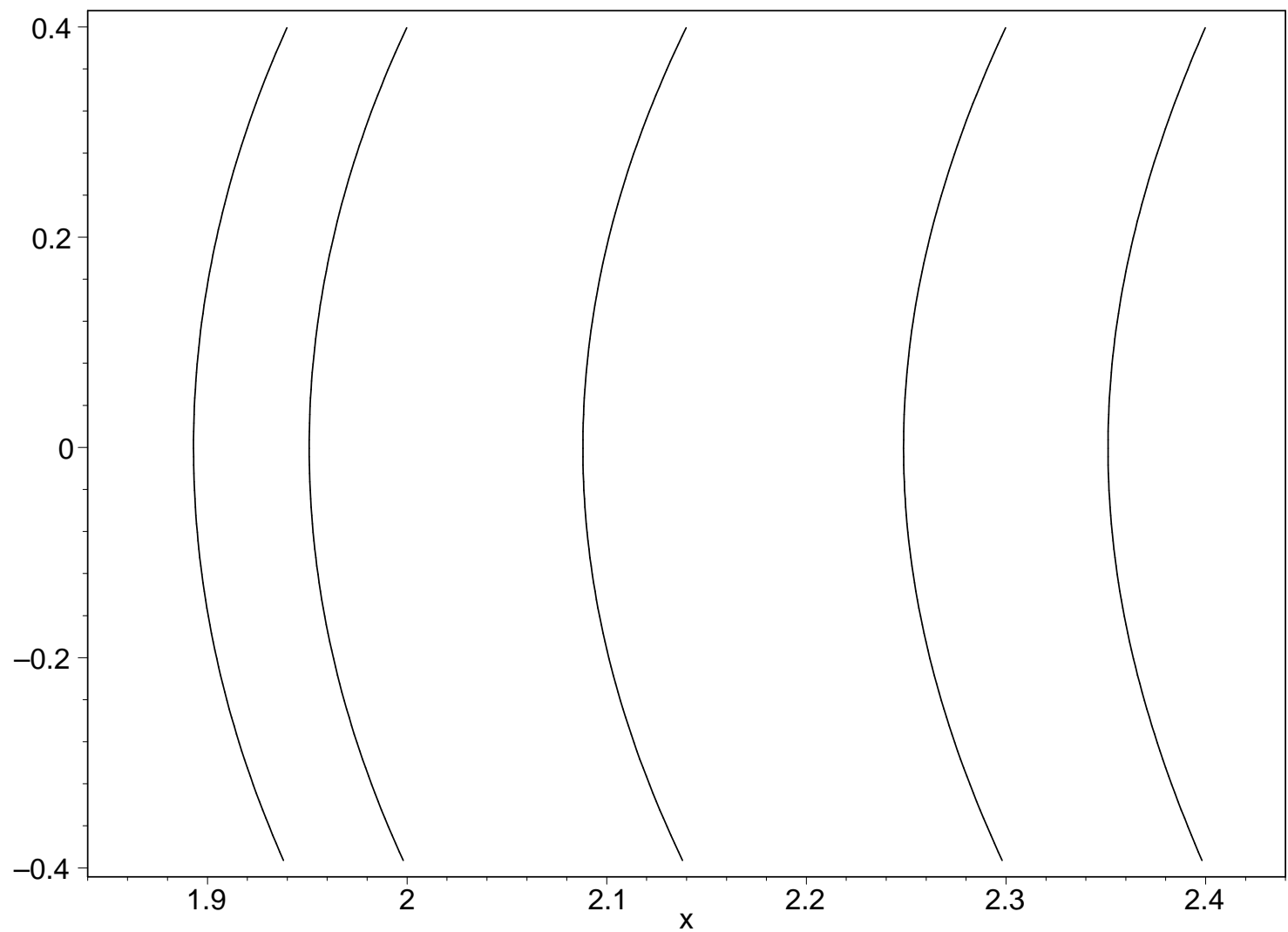
\end{document}